\documentclass[superscriptaddress,aps,prd,onecolumn,showpacs,floatfix,preprintnumbers,amsmath,amssymb,nofootinbib,groupedaddress]{revtex4}
\pdfoutput=1

\input epsf
\usepackage[pdftex]{graphicx}
\usepackage{graphicx}
\usepackage{color}
\usepackage{mathtools}

\usepackage[utf8]{inputenc}
\usepackage[T1]{fontenc}

\def\VEV#1{\left\langle #1 \right\rangle}

\newcommand{\M}{M_{pl}}

    \newcommand{\be}{\begin{equation}}
  \newcommand{\ee}{\end{equation}}
    \newcommand{\ba}{\begin{align}}
  \newcommand{\ea}{\end{align}}

\newcommand{\fsky}{f_{\rm sky}}

\definecolor{darkred}{rgb}{0.8,0,0}

\begin{document}

\title{Towards a measurement of the spectral runnings}

\author{Julian B. Mu\~noz\footnote{Electronic address: \tt julianmunoz@jhu.edu}
} 
\affiliation{Department of Physics and Astronomy, Johns
	Hopkins University, 3400 N.\ Charles St., Baltimore, MD 21218}
\author{Ely D. Kovetz}
\affiliation{Department of Physics and Astronomy, Johns
	Hopkins University, 3400 N.\ Charles St., Baltimore, MD 21218}
\author{Alvise Raccanelli$^\dagger$\let\thefootnote\relax\footnote{$^\dagger$Marie Sk\l{}odowska-Curie fellow}
} 
\affiliation{Department of Physics and Astronomy, Johns
	Hopkins University, 3400 N.\ Charles St., Baltimore, MD 21218}
\affiliation{Institut de Ci\`encies del Cosmos (ICCUB),Universitat de Barcelona (IEEC-UB), Mart\'{i} Franqu\`es 1, E08028 Barcelona, Spain }
\author{Marc Kamionkowski}
\affiliation{Department of Physics and Astronomy, Johns
	Hopkins University, 3400 N.\ Charles St., Baltimore, MD 21218}
\author{Joseph Silk}
\affiliation{Department of Physics and Astronomy, Johns
	Hopkins University, 3400 N.\ Charles St., Baltimore, MD 21218}
\affiliation{Institut d'Astrophysique de Paris, UMR 7095, CNRS, UPMC Univ. Paris VI, 98 bis Boulevard Arago, 75014 Paris,
	France}
\affiliation{BIPAC, Department of Physics, University of Oxford, Keble Road, Oxford OX1 3RH, UK
}

\date{\today}

\begin{abstract}

Single-field slow-roll inflation predicts a nearly scale-free power spectrum of perturbations,
as observed at the scales accessible to current cosmological experiments.
This spectrum is slightly red, showing a tilt $(1-n_s)\sim 0.04$.
A direct consequence of this tilt are
nonvanishing runnings $\alpha_s=\mathrm d n_s/\mathrm d\log k$, 
and $\beta_s=\mathrm d\alpha_s/\mathrm d\log k$, which in the minimal inflationary scenario
should reach absolute values of $10^{-3}$ and $10^{-5}$, respectively.
In this work we calculate how well future surveys can measure these two runnings. 
We consider a Stage-4 (S4) CMB experiment and show that it will be able to detect significant deviations from the inflationary prediction for $\alpha_s$, although not for $\beta_s$. 
Adding to the S4 CMB experiment the information from a WFIRST-like or a DESI-like survey improves the sensitivity to the runnings by $\sim$ 20\%, and 30\%, respectively. 
A spectroscopic survey with a billion objects, such as the SKA, will add enough information to the S4 measurements to allow a detection of 
$\alpha_s=10^{-3}$, required to probe the single-field slow-roll inflationary paradigm. We show that 
only a very-futuristic interferometer targeting the dark ages will be capable of measuring the minimal inflationary prediction for $\beta_s$.
The results of other probes, such as a stochastic background of gravitational waves observable by LIGO, the Ly-$\alpha$ forest, and spectral distortions, are shown for comparison.
Finally, we study the claims that large values of $\beta_s$, if extrapolated to
the smallest scales, can produce primordial black holes of tens of solar masses, which we show to
be easily testable by the S4 CMB experiment.
\end{abstract}

\maketitle

\section{Introduction}

The standard model of cosmology, or $\Lambda$CDM, has had great success 
explaining the cosmological observables within our reach \cite{1411.1074,1502.01589}.
In this model the universe is flat, homogeneous, and has perturbations characterized by an almost-scale-invariant power spectrum, arising during inflation \cite{Bardeen:1983qw,Mukhanov:1985rz,astro-ph/0504097}.
This primordial power spectrum creates overdensities that we can observe through temperature anisotropies in the cosmic microwave background (CMB) \cite{astro-ph/9609105}, through brightness fluctuations in the 21-cm hydrogen line \cite{astro-ph/0010468,astro-ph/0312134}, 
and with the cosmological large-scale structure, once these perturbations grow nonlinear \cite{astro-ph/9506072}.  

We parametrize deviations from perfect scale invariance by a few variables, 
which capture the change in the shape of the power spectrum at some {\it pivot} scale $k_*$. 
The first of these numbers is the scalar {\it tilt} $(1-n_s)$, which 
expresses a offset in the power-law index, and is measured to be more than 5
standard deviations smaller than zero \cite{1502.01589}, making 
the primordial scalar spectrum slightly red.
In the same spirit, higher derivatives, or {\it runnings}, of the power spectrum can be measured from cosmological data \cite{astro-ph/9504071}.
The first of these quantities is the scalar {\it running} $\alpha_s=\mathrm d n_s/\mathrm d \log k$, presently consistent with zero \cite{1502.01589}. However,
current Planck data seems to prefer a non-zero {\it second running} $\beta_s\equiv\mathrm d \alpha_s/\mathrm d \log k = 0.02\pm 0.01$, albeit only at 2$-\sigma$ \cite{1605.00209}.

In single-field slow-roll inflation scale invariance is predicted to extend over a vast range of scales \cite{0907.5424,1502.02114}.
However, we only have access to a small range of wavenumbers around the CMB pivot scale $k_*=0.05$ Mpc$^{-1}$.
The amplitude $A_s$ of the (scalar) power spectrum and its tilt $n_s$ give us information
about the first two derivatives of the inflaton potential 
when this scale $k_*$ exited the horizon during inflation.
Higher-order derivatives of this potential produce non-zero runnings,
which for slow-roll inflation generically have values $\alpha_s\sim(1-n_s)^2$ and 
$\beta_s\sim(1-n_s)^3$, beyond the reach of present-day cosmological experiments \cite{1007.3748}.
In this paper we will explore how well next-generation cosmological experiments, such as
the proposed Stage-4 (S4) CMB experiment, various galaxy surveys, and different 21-cm interferometers, can measure these numbers.
A detection of $\alpha_s$, or $\beta_s$, would
enable us to distinguish between inflationary models with otherwise equal predictions,
and shed light onto the scalar
power spectrum over a wider $k$ range. 

In the absence of any salient features in the power 
spectrum, such as small-scale non gaussianities, 
the power in the smallest scales will be determined by the runnings of the scalar amplitude.
This is of particular importance for primordial-black-hole (PBH) production in the early universe, 
where a significant increase in power is required at the scale corresponding to the 
PBH mass, which is of order $k \sim 10^5$ Mpc$^{-1}$ for solar-mass PBHs \cite{astro-ph/9901268,astro-ph/0511743}.
It has been argued that a value of the second running $\beta_s = 0.03$,
within 1$-\sigma$ of Planck results,
can generate fluctuations leading to the formation of $30\, M_{\odot}$ primordial black holes if extrapolated to the smallest scales \cite{1607.06077}, which could make up the dark matter \cite{1603.00464}.
 
We will show that the S4 CMB experiment will determine whether $\beta_s$ is
high enough to produce solar-mass PBHs, although it will not reach enough sensitivity in either $\alpha_s$ or $\beta_s$ to accurately test the slow-roll inflationary prediction.
Adding galaxy-clustering measurements to the S4 CMB will enhance these measurements,
with the improvement being $\sim 20\%$ for a WFIRST-{\it like}, and $\sim 30\%$ for a DESI-{\it like} survey, which will allow us to measure significant departures from single-field slow-roll inflation. 
Moreover, a billion-object survey, such as the SKA, will add enough information to the S4 experiment to be able to measure the running $\alpha_s$ with $10^{-3}$ precision, enough to test the inflationary prediction.
However, we show that in order to reach the sensitivity required for a measurement of $\beta_s\sim 10^{-5}$, a dark-ages interferometer, with a baseline of $\sim 300$ km, will be required.

This paper is structured as follows. In Sec.~\ref{sect:motivation} we 
will explore what can the runnings teach us about inflationary dynamics and PBH production,
as well as review the present bounds on $\alpha_s$ and $\beta_s$.
Later, in Sec.~\ref{sect:future} we will forecast the constraints on
the power spectrum amplitude, tilt, and runnings from different types of experiment. 
We conclude in Sec.~\ref{sect:conclusions}.

\section{Motivation and current constraints}
\label{sect:motivation}

During inflation, quantum fluctuations generate an almost-scale-invariant power spectrum of 
fluctuations. The scalar perturbations $\zeta_{\bf k}$ thus have a two-point function 
given by
\be
\VEV{\zeta_{\bf k}^{}\zeta_{\bf k'}^*} = P_{\zeta}(k) (2\pi)^3 \delta_D(\bf k+k'),
\ee
where $P_\zeta(k)$ is the scalar power spectrum, for which 
we can define an amplitude as
\ba
\log \Delta^2_s(k) &\equiv \log \left [\dfrac{k^3}{2\pi^2} P_\zeta(k) \right] = \log A_s + (n_s-1) \log\left(\dfrac k {k_*}\right) \nonumber \\ &+ \frac 1 2 \alpha_s \log^2\left(\dfrac k {k_*}\right)+ \frac 1 6 \beta_s \log^3\left(\dfrac k {k_*}\right),
\label{eq:Delta2}
\end{align}
where $A_s$ is the scalar amplitude, $n_s$ is the scalar tilt, and $\alpha_s$ and $\beta_s$ are the running
and the second running, respectively. At the pivot scale of $k_*=0.05$ Mpc$^{-1}$, Planck has measured a scalar amplitude $A_s = 2.196 \times 10^{-9}$, with tilt $n_s=0.9655$ \cite{1502.02114}. We will take these values, with $\alpha_s=\beta_s=0$, as our baseline. 
We will also assume as fiducial $\Lambda$CDM parameters $\omega_b=0.02222$, $\omega_c=0.1197$, $\tau = 0.06$, and $H_0 = 67.5$ km/s.

\begin{figure}[htbp!]
	\includegraphics[width=0.6\textwidth]{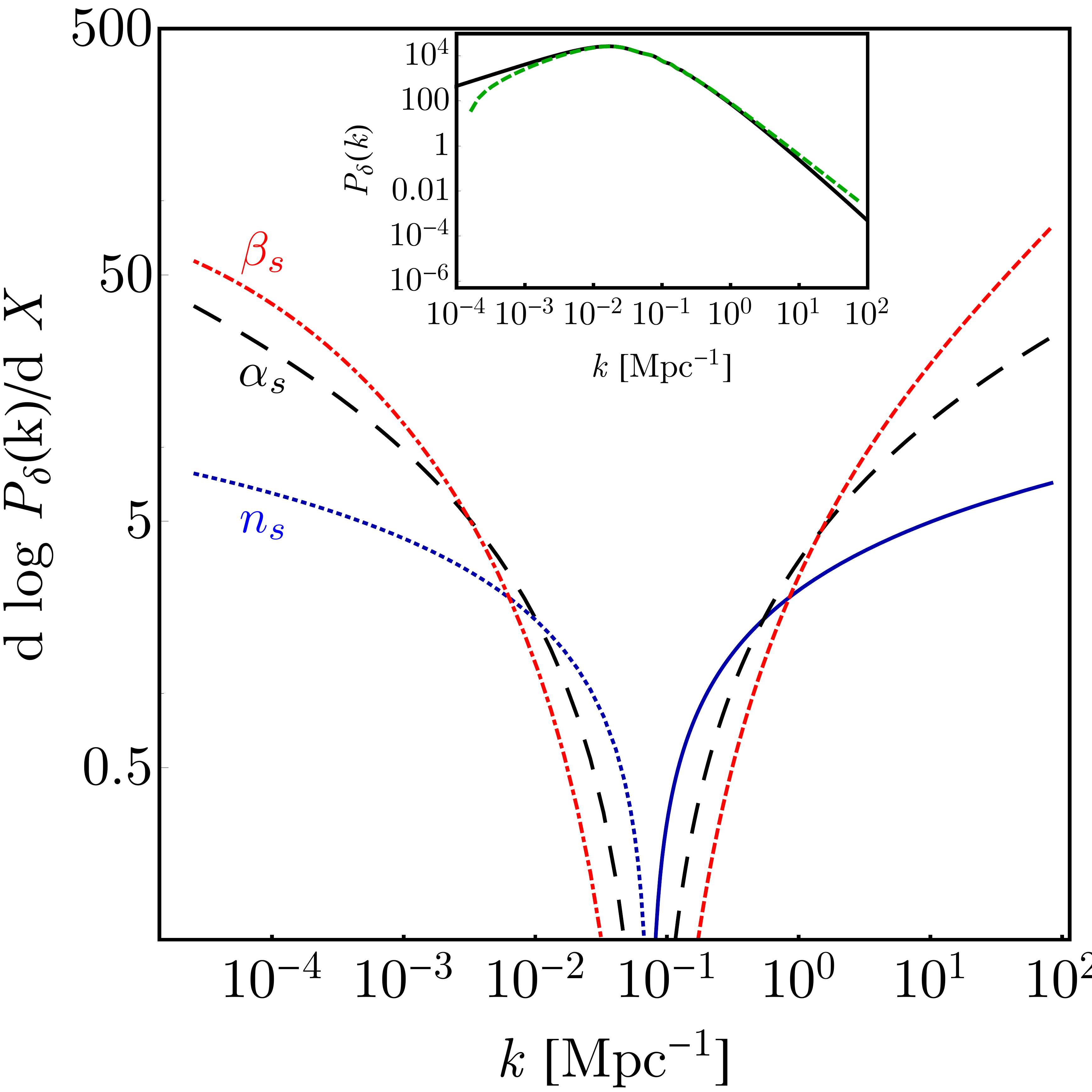}
	\caption{Logarithmic derivatives of the power spectrum $P_{\delta}(k)$, as a function of the wavenumber $k$ in Mpc$^{-1}$, with respect to the scalar tilt $n_s$ in blue line (solid for positive and dotted for negative values), to the running $\alpha_s$ in black long-dashed line, and the running of the running $\beta_s$ in red line (dashed for positive and dash-dotted for negative values), at the $\Lambda$CDM best-fit values. The matter power spectrum $P_{\delta}(k)$ is shown in the top center for comparison, where the case with no runnings corresponds to the  solid black line, and the case with $\beta_s=0.03$$-$the highest running allowed by Planck at 68\% C.L.$-$to the green-dashed line.}
	\label{fig:dlogP}
\end{figure}

The primordial perturbations $\zeta$, generated during inflation, create matter overdensities $\delta\equiv \rho/\bar \rho -1$, where $\rho$ is the energy density and $\bar{\rho}$ its spatial average.
These matter perturbations source the temperature fluctuations in the CMB and later on grow to seed
the large-scale structure of the universe.
In linear theory, matter and primordial perturbations
are related to each other through a transfer function 
$\mathcal T(k)$, which can be calculated with a Bolztmann code like {\tt class} \cite{1104.2933} or  {\tt CAMB} \cite{astro-ph/9911177},
so that the matter power spectrum is
\be
P_{\delta} (k) = \mathcal T^2(k) P_{\zeta}(k).
\ee
We show the logarithmic derivative of the matter power spectrum with respect to the 
tilt $n_s$, the running $\alpha_s$, and the second running $\beta_s$, at our fiducial values in Fig.~\ref{fig:dlogP}.
From this Figure it is clear that scales away 
from the pivot scale $k_*=0.05$ Mpc$^{-1}$ change the most when higher-order runnings are introduced.
Note that the logarithmic derivative with respect to the scalar amplitude $A_s$ would just be a horizontal line in this plot.
We also show the regular matter power spectrum $P_{\delta}$ for comparison, both without any runnings and with $\beta_s=0.03$, as argued in Ref.~\cite{1607.06077} to be enough to produce PBHs.

\subsection{Single-Field Slow-Roll Inflation}

Let us now briefly review the dynamics of single-field slow-roll (SFSR) inflation, and how the runnings change it.
In the case of inflation being driven by a single field $\phi$ under a potential $V(\phi)$,
the amplitude $A_s$ and tilt $n_s$ of the scalar power spectrum
are determined by a combination of $V'(\phi_*)$ and $V''(\phi_*)$ \cite{1502.02114}, 
where $\phi_*$ is the value of the field at the pivot scale, 
and $'$ denotes a derivative
with respect to $\phi$. 
The absolute magnitude of $V(\phi_*)$ is related to the tensor-perturbation amplitude,
or alternatively to its ratio $r$ to the scalar amplitude \cite{astro-ph/0601276,1111.3040}.
The uncertainty in the reheating phase at the end of inflation, however,
hampers a unique determination of the shape of the inflaton potential
from $A_s$, $n_s$, and $r$ \cite{1001.2600,1404.6704,1410.3808,1412.0656}. 
The inclusion of an additional observable, such as the scalar running $\alpha_s$,
may help alleviate these uncertainties, as it provides information about the $V'''(\phi_*)$ term \cite{astro-ph/9408015,1505.00968}.
A similar argument applies to the second running $\beta_s$ with the fourth derivative of the potential.
Unfortunately, in single-field slow-roll inflation these runnings are expected
to be rather small.
To illustrate why, let us define the slow-roll parameters
\ba
\epsilon &= \dfrac {\M^2} {2} \left( \dfrac{V'}{V} \right)^2,\nonumber\\
\eta &= \M^2 \dfrac{V''}{ V},\nonumber\\
\xi^2 &=  \M^4 \dfrac{V'V'''}{V^2},\nonumber\\
\sigma^3 &= \M^6 \dfrac{V'^2 V^{(4)}}{V^3},
\end{align}
where $\M=(8\pi G)^{-1/2}\approx 2.4\times 10^{18}$ GeV$/$c$^2$ is the reduced Planck mass, and we have defined the third- and fourth-order slow-roll parameters, $\xi$ and $\sigma$ respectively, to be of the same order as $\epsilon$ and $\eta$ in SFSR inflation.
In this case, both the scalar and tensor indices (denoted by $n_t$, $\alpha_t$, and $\beta_t$) can be found to first non-vanishing order in the slow-roll parameters as~\cite{Stewart:1993bc,1303.1688}
\ba
& & r&= 16\epsilon,\nonumber\\
1-n_s &= 2\eta-6\epsilon  & n_t&=-2\epsilon, \nonumber\\
\alpha_s &= -2\xi^2+16\eta\epsilon-24\epsilon^2 & \alpha_t &= 4\eta\epsilon - 8 \epsilon^2,\nonumber\\
\beta_s &=  2\sigma^3 +2\xi^2(\eta-12\epsilon) - 32\epsilon(\eta^2-6\eta\epsilon+6\epsilon^2) & \beta_t &= -4\epsilon\xi^2-8\epsilon\left(\eta^2-7\epsilon\eta+8\epsilon^2\right),
\label{eq:SFSRruns}
\end{align}
so for $\sigma\sim\xi\sim\eta$ the prediction of single-field slow-roll inflation is 
$\alpha_s =O[(n_s-1)^2]\sim 10^{-3}$, and $\beta_s =O[(n_s-1)^3]\sim 10^{-5}$,
as long as the potential does not experience a sudden change near CMB scales \cite{astro-ph/9504071,astro-ph/0206032,1011.3988,1303.3925,1309.1285,1402.2059,1610.09362}.
In the same way that large local non-gaussianities would rule out single-field slow-roll inflation \cite{astro-ph/0210603,astro-ph/0407059},
a running $\alpha_s$ much larger than $\sim\,10^{-3}$ would also imply 
a more complex model of inflation.
A similar argument holds with $\beta_s$, although with a 
$10^{-5}$ amplitude. 
Furthermore, it has been argued that very-large positive values of $\beta_s \gtrsim 10^{-2}$ are not possible within any SFSR model, as they would force the inflationary era to end before the largest observable 
scales exited the horizon \cite{astro-ph/0206032,1209.2024,1605.00209}. 
Thus, PBH production with a value of the second running $\beta_s = 0.03$
requires additional degrees of freedom during inflation.
In general, reaching a precision of $\sigma(\beta_s)\lesssim 10^{-2}$ 
will provide a useful test of the single-field inflationary paradigm. 

\subsubsection{Gravitational Waves}

From Eq.~\eqref{eq:SFSRruns} it is clear that the tensor running $\alpha_t$ is a function solely of $\epsilon$ and $\eta$, which can be determined from the observables $n_s$ and $r$. Moreover, the tensor second running $\beta_t$ also depends on the third-order slow-roll parameter $\xi$, which can then be inferred from the scalar running $\alpha_s$. This allows us to write the tensor indices as a function of the scalar ones, plus $r$, in the following form
\ba
n_t &= -\dfrac r 8 ,\nonumber\\
\alpha_t &= \dfrac r {64} \left[ r + 8 (n_s-1) \right],\nonumber\\
\beta_t &= -\dfrac{r}{256} \left[ -32 \alpha_s + r^2 +32 (n_s-1)^2+12 (n_s-1)r \right].
\label{eq:SFSRrunst}
\end{align}
For any observed value of $r$ and $n_s$---measured at CMB scales---we can plug in the right-hand side of Eq.~\eqref{eq:SFSRrunst} and find $n_t$ and $\alpha_t$, which lets $\beta_t$ vary as a function of $\alpha_s$ alone. 
Then, using the equivalent of Eq.~\eqref{eq:Delta2} for tensor perturbations,
we can predict the gravitational-wave amplitude
observed by Earth-based interferometers (see e.g.~Ref.~\cite{Smith:2005mm}).
Given the large lever-arm between these observations, performed at $k = O( 10^{15})$ Mpc$^{-1}$, and those at CMB scales, direct-detection searches of a gravitational-wave background can probe large and positive values of $\beta_t$, as a function of $r$. Therefore, they also indirectly constrain the scalar running $\alpha_s$.
For instance, the first run (O1) of LIGO placed a constraint on the gravitational-wave energy density of $\Omega_{\rm GW}\lesssim 5\times10^{-7}$ at a frequency $f\approx30$ Hz \cite{TheLIGOScientific:2016wyq}. This corresponds to a constrain on $\beta_t \lesssim 2\times 10^{-3}$, for $r>10^{-3}$, or, assuming SFSR and using Eq.~\eqref{eq:SFSRrunst}, to $\alpha_s < 0.014/r$. Moreover, the O2 and O5 runs of LIGO will improve the constraints on the stochastic gravitational-wave background to $\Omega_{\rm GW}\lesssim 5\times10^{-8}$ and $\Omega_{\rm GW}\lesssim 5\times10^{-9}$, respectively. This will allow LIGO to probe $\alpha_s$ larger than $0.012/r$ after O2, and $0.009/r$ with O5 data.

\subsection{PBHs}

A large positive value of the second running $\beta_s$ has consequences for 
primordial-black-hole (PBH) formation.  
There has been recent interest in PBHs as a dark matter candidate \cite{1607.06077,CarrHawking,Carr,Mesz,1501.07565,1603.00464,Clesse:2016vqa},
since they could explain some of the gravitational-wave events 
observed by the LIGO collaboration \cite{1602.03837}.

If they are to be the dark matter, PBHs could have formed in the primordial universe from very-dense pockets
of plasma that collapsed under their own gravitational pull.
The scales in which stellar-mass PBHs were formed are orders of magnitude
beyond the reach of any cosmological observable. 
However, if the inflationary dynamics were fully determined by a single field, 
one could extract information about the potential $V(\phi)$ at the smallest
scales from $V(\phi_*)$ at the pivot scale (and its derivatives) by extrapolation.

The formation process of the PBHs is poorly understood \cite{astro-ph/9901268},
so we will not attempt to model it, and
instead we will assume that PBHs form at the scale at which $\Delta_s^2(k)$ becomes of order unity. 
It is clear that any positive running, if not compensated by a negative running of
higher order, will create enough power in some small-enough scale to have $\Delta_s^2(k)=1$.
Nonetheless, we will require that the mass of the formed PBHs is larger than $\sim 10^{15}$ gr, 
to prevent PBH evaporation before $z=0$, 
which sets a limit on the smallest scale where PBHs can form of
of $k_{\rm pbh} = 10^{15}$ Mpc$^{-1}$. 
We compute the $\alpha_s-\beta_s$ range in parameter space that produces enough power 
in scales $k<k_{\rm pbh}$ to generate PBHs, from Eq.~\eqref{eq:Delta2}, and show it in Fig.~\ref{fig:total}, along with the constraints from future experiments.
In order to produce PBHs of $\sim 30 \, M_{\odot}$, as suggested in Ref.~\cite{1603.00464} to be the dark matter, the relevant scale is $k\sim 10^5$ Mpc$^{-1}$, forcing the second running to be as large as $\beta_s \approx 0.03$, which will be tested at high significance by the S4 CMB experiment, as well as a DESI-like galaxy survey.

Notice that, even though there is no reason for the expansion in Eq.~\eqref{eq:Delta2}
to be truncated at second order in $\log(k/k_*)$, the 
next term in the series would be a factor $4 \beta_s/[ \gamma_s \times \log(k/k_*)]$ smaller,
where $\gamma_s\equiv \mathrm d \beta_s/\mathrm d\log k$ is the next-order running. 
For $\gamma_s\sim (n_s-1)\times \beta_s$, and for all scales $k<k_{\rm pbh}$ that form non-evaporating PBHs, the $\gamma_s$ term would be at least $\sim 3$ smaller 
than the $\beta_s$ term, making it subdominant in our order-of-magnitude estimate.
In any case we note that the values of $\beta_s$ required for PBH formation are $\sim 10^{-3}$,
two orders of magnitude larger than the standard slow-roll prediction \cite{astro-ph/9504071}.

\subsection{Current constraints}

We will now review previous constraints on the runnings and their
proposed improvements.

\subsubsection{Spectral Distortions}

Changes in the black-body spectrum of the CMB can arise due to early energy injection,
creating spectral distortions. If observed, these spectral distortions
can probe very-small scales during the post-thermalization epoch of the early universe. 
The FIRAS experiment showed that the CMB behaves as a black body, constraining
the spectral distortions to be smaller than roughly one part in $10^5$ \cite{astro-ph/9605054}. 
This null detection can only be used to constrain very positive values of the runnings, as
negative values would create even less power 
at small scales, and hence less spectral distortions.

The upcoming PIXIE experiment will detect spectral distortions down
to one part in $10^8$, within the range of values expected to be produced by $\Lambda$CDM \cite{1105.2044,1603.02496}. 
An improved version of PIXIE would even allow us to measure the running with accuracy $\sigma(\alpha_s) = 10^{-2}$ if we observe spectral distortions \cite{1209.2024,1602.05578}, and if there is not enough power at small scales to produce them we will be able to infer a negative running $\alpha_s < 0$.
Similarly, PIXIE will enable us to measure 
the second running with accuracy $\sigma(\beta_s) = 8\times10^{-3}$ \cite{1605.00209}.
As we will see in Sec.~\ref{sect:S4}, however, these constraints are less stringent than the ones expected from S4 CMB.

\subsubsection{Ly-$\alpha$ Forest}

Observations at smaller scales than the CMB provide a longer lever arm to measure the running of the
power spectrum. Ly-$\alpha$ forest data can
reach wavenumbers $k\sim $ few $\times$ Mpc$^{-1}$ \cite{1410.7244}, a factor of $\sim10$ larger than the CMB.
This was used in Ref.~\cite{1506.05976} to fit for the running $\alpha_s$ with joint 
Ly-$\alpha$ and Planck data, 
where it was found that $\alpha_s = -0.0152^{+0.0050}_{-0.0045}$, roughly 3-$\sigma$ below zero, and an order of magnitude larger than the single-field slow-roll prediction.
In practice, however, different systematics, such as baryonic physics \cite{1207.6567}, hinder our ability to accurately map a set of primordial parameters to an observed Ly-$\alpha$ power spectrum \cite{Bird:2010mp}.
In particular, the scalar tilt $n_s$ measured at CMB and Ly-$\alpha$ scales appears to be at tension, 
which is argued to drive the non-zero value of the running~\cite{1506.05976}.

\section{Future Constraints}
\label{sect:future}

In this section we will forecast, via Fisher analysis, how well different observables can measure the runnings. 
We will first calculate the limits that can be reached by a S4 CMB experiment in Sec.~\ref{sect:S4}. 
We will then study the forecasts from power-spectrum measurements with forthcoming galaxy surveys, centering our analysis on the planned DESI\footnote{http://desi.lbl.gov/}, WFIRST\footnote{http://wfirst.gsfc.nasa.gov/}, and the Square Kilometre Array (SKA\footnote{https://www.skatelescope.org/}) in Sec.~\ref{sect:LSS}.
Afterwards, we will perform an order-of-magnitude forecast with different 21-cm experiments, focusing on the proposed FFTT in Sec.~\ref{sect:21cm}.
Finally, we will combine the results to find the ultimate constraints in Sec.~\ref{sect:combined}.

\subsection{CMB}
\label{sect:S4}

The proposed S4 CMB experiment will be able to map
modes up to $\ell\sim 5000$, both in temperature and polarization.
We will study the level of precision that this S4 CMB experiment can reach for
the running $\alpha_s$, and the second running $\beta_s$.

\subsubsection{Formalism}

The CMB power spectra can be written as
\be
C_{\ell}^{XY} = (4\pi)^2 \int \mathrm d k\, k^2{\cal T}_\ell^X(k) {\cal T}_\ell^Y (k) P_\zeta(k),
\ee
where the indices $X,Y =\{T,E,$ or $d\}$ stand for temperature, E-mode polarization, and lensing potential respectively, and
${\cal T}_\ell^X$ are their transfer functions \cite{astro-ph/9611125,astro-ph/9603033,Kamionkowski:1996zd,astro-ph/9609169}. These ${\cal T}_\ell^X$
do not depend on the primordial power spectrum, so the runnings only affect the $C_{\ell}$
through the change in $P_\zeta(k)$.

To forecast the errors in a set of parameters $\theta_i$ we 
define the Fisher matrix as \cite{Jungman:1995av,Jungman:1995bz,1402.4108}
\be
F_{ij} = \sum_{\ell} \dfrac{2\ell+1}{2} \fsky {\rm Tr}\left [ \mathbf C_\ell^{-1} \dfrac{\partial\mathbf C_\ell}{\partial \theta_i} \mathbf C_\ell^{-1}  \dfrac{\partial\mathbf C_\ell}{\partial \theta_j} \right],
\ee
where $f_{\rm sky}$ is the sky-fraction covered, and the covariance matrix, ignoring $E-d$ correlations is given by
\begin{equation}
\mathbf C_\ell =  \left ( 
\begin{tabular}{c c c}
$\tilde C_\ell^{TT}$ \quad & $ C_\ell^{TE}$ & $C_\ell^{Td} $ \\

$ C_\ell^{TE}$ \quad & $\tilde C_\ell^{EE}$ & $0$ \\

$ C_\ell^{Td}$ \quad & $0$ & $ \tilde C_\ell^{dd} $ \\
\end{tabular}
\right),
\end{equation}
and we have defined \cite{1403.5271,1511.04441}
\ba
\tilde C_\ell^{TT} &\equiv C_\ell^{TT} + N_\ell^{TT}, \nonumber \\ 
\tilde C_\ell^{EE} &\equiv C_\ell^{EE} + N_\ell^{EE} , \nonumber \\ 
\tilde C_\ell^{dd} &\equiv C_\ell^{dd} + N_\ell^{dd},
\end{align}
where $N_\ell^{XX}$ are the noise power spectra, given by
\ba
N_\ell^{TT} &= \Delta_T^2 \, e^{\ell (\ell+1) \sigma_b^2}, \nonumber \\ 
N_\ell^{EE} &= 2 \times N_\ell^{TT},
\end{align} 
where $\Delta_T$ is the temperature sensitivity, and $\sigma_b=\theta_{\rm FHWM}/\sqrt{8\log 2}$,
with the full-width-half-maximum $\theta_{\rm FHWM}^2$ given in radians. For the lensing noise $N_\ell^{dd}$ we follow the approach in Ref.~\cite{1010.0048}, where the E- and B-mode data is used to reconstruct the lensing power spectrum, $C_\ell^{dd}$, whose effect is then subtracted from the B-mode data. This allows us to iteratively compute the maximum delensing possible given the S4 polarization noise, and thus forecast the sensitivity to lensing modes \cite{Knox:2002pe,Kesden:2002ku,astro-ph/0301031,astro-ph/0306354}. 
For further details of this procedure the reader is encouraged to visit Section 7.4 in the CMB-S4 science book \cite{1610.02743}. For reference, the S4 lensing noise is predicted to be smaller than the signal for $\ell\lesssim 1000$.

The specifications we use for the S4 CMB experiment follow those of Ref.~\cite{1610.02743}, 
which has a sensitivity $\Delta_T = 1\,\mu$K-arcmin, with a resolution of 
$\theta_{\rm FWHM} = 3$ arcmin, over 40$\%$ of the sky. To that we add Planck
over an additional $20\%$ of the sky, and a prior on the optical depth of reionization of $\tau = 0.06 \pm 0.01$. The S4 experiment is expected to observe the $\ell$ range
between 30 and 5000 for polarization, although the highest modes will be noise-dominated;
and between $\ell =$ 30 and 3000 for temperature, as higher multipoles would be contaminated by foregrounds.
For Planck we take two bands, corresponding to frequencies of 143 and 217 GHz, respectively, with noises $\Delta_T = \{43, 66\}\,\mu$K-arcmin, and $\Delta_E = \{81,134\}\,\mu$K-arcmin, a resolution of $\theta_{\rm FWHM} = \{7,5\}$ arcmin and we do not include lensing data.
	
The parameters $\theta_i$
that we will forecast in this section are the six $\Lambda$CDM parameters ($\omega_b$, $\omega_c$, 
$n_s$, $A_s$, $\tau$, and $H_0$), plus the running $\alpha_s$, and the second running $\beta_s$. 

\subsubsection{Results}

\begin{figure}[hbtp!]
		\includegraphics[width=0.99\textwidth]{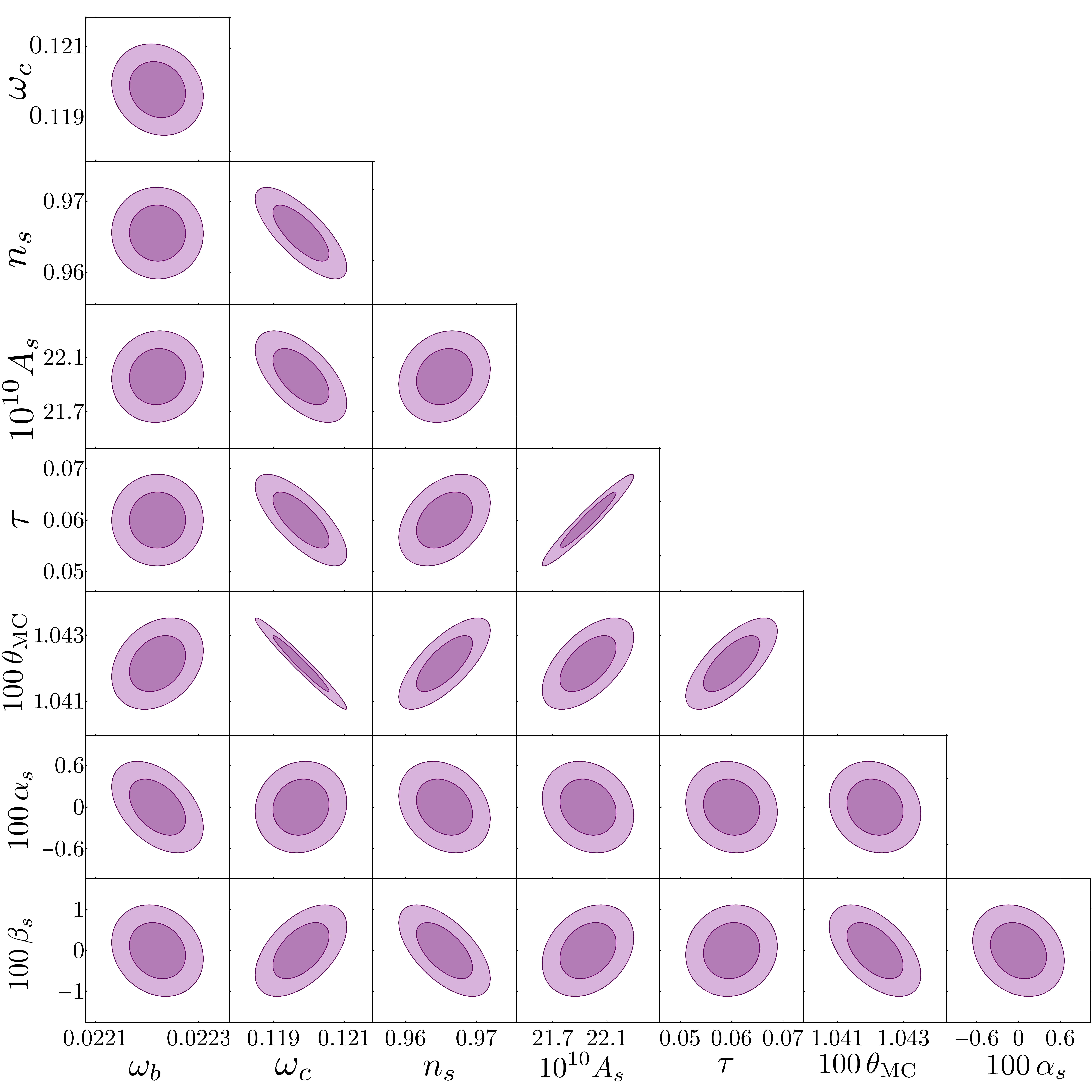}
		\caption{Confidence ellipses for the $\Lambda$CDM parameters and $\alpha_s$ and $\beta_s$, for the S4 CMB experiment. In darker purple we show the 68\% C.L. region, and in lighter purple the 95\% C.L. region.}
		\label{fig:S4}
	\end{figure}

We show in Fig.~\ref{fig:S4} the confidence ellipses between $\alpha_s$, $\beta_s$, 
and the six $\Lambda$CDM parameters. 
Both runnings are mainly degenerate with $\omega_b$, $n_s$, and $A_s$. 
Moreover, $\alpha_s$ and $\beta_s$ are also correlated,
which is to be expected, since both $\alpha_s$ and $\beta_s$ increase the power at $k>k_*$, or $\ell\gtrsim500$, 
where a great part of the CMB information comes from.
We show the forecast uncertainties for these parameters in Table~\ref{tab:S4}.
In this table we show the minimum $\alpha_s$ that could be measured by a S4 CMB experiment (marginalizing over $\Lambda$CDM parameters but setting $\beta_s=0$), which is $\sigma(\alpha_s)=0.0025$, enough to 
detect significant departures from slow-roll single-field inflation, 
albeit not sufficient to detect the slow-roll prediction $\alpha_s \approx 10^{-3}$. 
Meanwhile, the 1$-\sigma$ C.L. on 
$\beta_s$ (marginalizing over $\Lambda$CDM$+\alpha_s$) will be $\sigma(\beta_s)=0.0045$. 
This will shed light on the claimed detection of a non-zero second running $\beta_s = 0.02\pm0.01$ \cite{1605.00209},
and will have the power to test whether primordial black holes with tens of solar masses are formed from a positive second running.  
From these results we find that the S4 experiment alone will not suffice to determine the dynamics of inflation.
Similar results have been forecasted for the proposed COrE satellite \cite{Escudero:2015wba}.

As an additional test we have studied the correlation between both the running $\alpha_s$, and the second running $\beta_s$, and the neutrino mass $m_\nu$ for the S4 experiment. We find a correlation, defined as $r_{ij} \equiv C_{ij}/\sqrt{C_{ii}C_{jj}}$, with $C\equiv F^{-1}$, of $r_{\alpha_s,m_\nu}=-0.24$ between the neutrino mass and the running, and $r_{\beta_s,m_\nu}=-0.23$ between the neutrino mass and the second running. Marginalizing over $m_\nu$ in addition to the other $\Lambda$CDM parameters worsens the sensitivity of the S4 CMB experiment to the runnings at percent level, which justifies neglecting the effects of the mass of the neutrinos in our forecasts.

	\begin{table*}[hbtp!]
		\begin{tabular}{|l | c | c | c | c | c | c | c | c |  }
			\hline
			Model & $\sigma\left(\omega_b\right)$ & $\sigma\left(\omega_c\right)$ & $\sigma\left(n_s\right)$ & $\sigma\left(A_s\right)$ & $\sigma\left(\tau\right)$ & $\sigma\left(H_0\right)$ &  $\sigma\left(\alpha_s\right)$&  $\sigma\left(\beta_s\right)$   \\             
			\hline
			\hline
			$\Lambda$CDM+$\alpha_s$ &  3.4 $\times 10^{-5}$ & 6.0 $\times 10^{-4}$ & 2.2 $\times 10^{-3}$ & 	2.1 $\times 10^{-11}$ & 0.0055 & 0.23 & 
			  0.0025 & $-$ \\
				$\Lambda$CDM+$\alpha_s$+$\beta_s$ & 3.5 $\times 10^{-5}$ & 7.0 $\times 10^{-4}$ & 2.7 $\times 10^{-3}$ & 	2.1 $\times 10^{-11}$ & 0.0056 & 0.27 & 
			 0.0026 & 0.0045 \\
			\hline			
		\end{tabular}
		\caption{$1-\sigma$ C.L. forecast for the S4 CMB experiment. We consider the six $\Lambda$CDM parameters and $\alpha_s$ in the first row, and we add the second running $\beta_s$ in the last row.}
		\label{tab:S4}
	\end{table*}

\subsection{Galaxy Clustering}
\label{sect:LSS}

Galaxy clustering has been used for measurements of a variety of cosmological parameters (see e.g.~\cite{Tegmark:2003ud, SDSSDR7, BOSSDR9, Raccanelli:growth, BOSSDR12, BOSSDR12LOS}).
In this section we will use the galaxy power spectrum to forecast the precision in measurements of the spectral index and its runnings, as defined in Eq.~\eqref{eq:Delta2}, for a variety of surveys. In order to have a good understanding of the regimes where galaxy clustering is more effective in providing these constraints, we consider a wide survey like DESI~\cite{Levi:2013gra} and a narrow but deep one such as the planned space telescope WFIRST~\cite{1503.03757}. We then explore what results could be obtained by a wide and deep galaxy survey such as the Square Kilometre Array (SKA) HI spectroscopic galaxy survey~\cite{1501.04035}.
For the specifications of the surveys we follow Ref.~\cite{Font-Ribera:2013rwa} for DESI, Ref.~\cite{1503.03757} for WFIRST, and for the SKA HI we extract our catalog from the S$^3$-SAX part of the $S^3$ simulation\footnote{http://s-cubed.physics.ox.ac.uk}, using a flux limit $S$ of 1 $\mu$Jy, which yields observations of over a billion objects over 30,000 deg$^2$ of the sky (see also Ref.~\cite{Santos} for more details about redshift distributions and biases of the SKA at different flux limits).

\subsubsection{Formalism}

The relationship between the real-space and the measured position of a galaxy is modified in the presence of peculiar velocities, giving rise to redshift-space distortions (RSDs). The matter overdensity in redshift space (labeled as $s$) can be linked to the real space one by using the RSD operator, defined (in the plane-parallel and linear approximations) as~\cite{Kaiser:1987qv}
\begin{equation}
\label{eq:kaiser}
\delta^s(k) = \left( 1+\beta \mu^2 \right) \delta^r(k) ,
\end{equation}
where $\delta^r(k)$ is the Fourier mode of the overdensities in real space, $\mu$ is the cosine of the angle with the line-of-sight, $\beta\equiv f(z)/b(z)$, where $b(z)$ is the bias, and $f(z)\equiv \mathrm d \log D/ \mathrm d \log a$ is the logarithmic derivative of the growth factor.
Therefore, the redshift-space galaxy power spectrum can be written as
\begin{align}
P^s_{\rm g}(k,\mu,z) = \left[ b(z) + f(z) \mu^2 \right]^2 P_{m}^r(k,z) + P_{\rm shot}(z) \, ,
\end{align}
where $P_{m}^r$ is the matter power spectrum in real space~\cite{Seo:2003pu} and the shot noise contribution is the inverse of the galaxy number density at that redshift, $P_{\rm shot}(z) \equiv \left[\bar{n}_g(z)\right]^{-1}$.

We limit our analysis to linear scales, but as a cautionary measure we include the Fingers-of-God effect~\cite{Jackson:1972} parametrized as ${\rm FoG} = e^{ -k^2 \mu^2 \sigma_v^2/H_{\rm 0}^2 }$, where $\sigma_v$ is the velocity dispersion and is modeled as in~\cite{Raccanelli:anifNL}.
A real data analysis will need to take into account a variety of additional corrections, such as wide-angle effects \cite{Szalay:1997cc, Raccanelli:2010hk, Raccanelli:2013gja}, and cosmic magnification~\cite{LoVerde:2006cj, Raccanelli:2011pu}; however none of these effects should be degenerate with modifications of the power spectrum spectral index or its runnings, so we will not include a detailed modeling of them.

Given the specifications of a survey, the Fisher analysis allows us to estimate the errors on the cosmological parameters around the fiducial values. We write the Fisher matrix for the power spectrum in the following way~\cite{astro-ph/9706198}
\begin{align}
\label{eq:FM}
F_{AB} &= \int_{z_{\rm min}}^{z_{\rm max}} \mathrm dz \int_{k_{\rm min}}^{k_{\rm max}}\mathrm dk \, k^2 \int_{-1}^{+1}\mathrm d\mu \frac{V_{\rm eff}(k,\mu,z)}{8\pi^2 \left[P_g^s(k,\mu,z)\right]^2}  \frac{\partial P_g^s(k,\mu,z)}{\partial \vartheta_A}\frac{\partial P_g^s(k,\mu,z)}{\partial \vartheta_B} B_{\rm nl} \, ,
\end{align}
where $\vartheta_{A}$ runs over $\{\omega_b,\omega_c,n_s,b,H_0,\alpha_s,\, \beta_s\}$, and the effective volume of the survey in the $z$-th redshift bin is 
\begin{equation}
V_{\rm eff}(k,\mu,z) = V_s \left[\frac{\bar{n}_g(z) P_g^s(k,\mu,z)}{1+\bar{n}_g(z) P_g^s(k,\mu,z)}\right]^2 \, ;
\end{equation}
$V_s$ is the volume of the survey, and $\bar{n}_g$ is the mean comoving number density of galaxies.
The $B_{\rm nl}$ term in Eq.~(\ref{eq:FM}) is given by
\begin{equation}
B_{\rm nl} = e^{-k^2\Sigma_{\perp}^2 -k^2 \mu^2 \left( \Sigma_{||}^2 -\Sigma_{\perp}^2 \right) } ,
\end{equation}
and accounts for non-linearities induced by the BAO peak~\cite{Seo:2003pu},
and $\Sigma_\bot=\Sigma_0D$, $\Sigma_{||}=\Sigma_0(1+f)D$, where $\Sigma_0$ is a
phenomenological constant describing the diffusion of the BAO peak due to nonlinear evolution. From N-body simulations its numerical value is 12.4 $\rm h^{-1} Mpc$ and depends weakly on $k$ and cosmological parameters~\cite{Eisenstein:2006nk, Padmanabhan:2012}.

Eq.~(\ref{eq:FM}) involves an integral over the wavenumber $k$; the largest scale that can be probed is determined by the geometry of the survey, $k_{\rm min} = 2\pi V_s^{-1/3}$, while we choose the smallest scale $k_{\rm max}$ to be the non-linear scale, computed as the wavenumber for which $\VEV{\delta^2} = \int_{k_{\rm min}}^{k_{\rm max}}dk \,k^2 P_\delta(k,z)/(2\pi^2)=0.6$, conservatively at the edge of the linear regime \cite{Giannantonio:2012}.

\subsubsection{Results}
In Table~\ref{tab:results_pk} we present forecasts for the precision of measurements of the spectral index and its runnings, for the different surveys described above, added to the Planck CMB experiment as defined in Section~\ref{sect:S4}.
We remain conservative by marginalizing over the bias $b$ before adding the information of
our galaxy surveys to Planck. Once we add the Fisher matrices of Planck and each survey we marginalize over the six $\Lambda$CDM parameters to forecast for $\alpha_s$, and over $\alpha_s$ as well for $\beta_s$. 
See for example Ref.~\cite{Ballardini:2016hpi} for a similar analysis.
From Table~\ref{tab:results_pk} it is clear that a wide survey, like DESI, will be more successful at measuring the runnings than a deep one, like WFIRST. Moreover, either of these surveys added to Planck will reach comparable sensitivity in $\alpha_s$ and $\beta_s$ to the S4 CMB experiment alone.
However, a more futuristic billion-object galaxy survey, such as the planned SKA, would yield better precision than S4 CMB.
Moreover, multi-tracer (MT) analyses~\cite{McDonald:2009, Seljak:2009}, where one can divide the galaxy catalog into subsamples of different populations, can potentially improve the constraints from galaxy surveys by a factor of a few, depending on the assumptions on the number of distinguishable populations and their bias (for some first analyses using the MT technique, see Refs.~\cite{GAMA:MT, Ferramacho:2014, dePutter:MT, Raccanelli:anifNL,Fonseca:2015laa,Alonso:2015sfa}).

\begin{table*}[hbtp!]
	{\small
		\begin{tabular}{| l | c | c |}
			\hline
			Experiment & $\sigma\left(\alpha_s\right)$ & $\sigma\left(\beta_s\right)$ \\
			 \hline\hline
			Planck CMB & 0.0053 & 0.0091 \\			
			Planck+WFIRST & 0.0026 & 0.0044 \\
			Planck+DESI & 0.0023 & 0.0042 \\
			Planck+SKA & 9.3 $\times 10^{-4}$ & 0.0020 \\
			 \hline
		\end{tabular}
	}
	\caption{68\% C.L. Uncertainties in the scalar running $\alpha_s$, and second running $\beta_s$ for the Planck experiment and different galaxy surveys.
		We marginalize over the six $\Lambda$CDM parameters (plus the bias amplitude $b$ for the galaxy surveys) when computing $\sigma(\alpha_s)$ and over $\alpha_s$ as well for $\sigma(\beta_s)$.
	}
	\label{tab:results_pk}
\end{table*}

\begin{figure*}[htbp!]
	\includegraphics[width=0.7\textwidth]{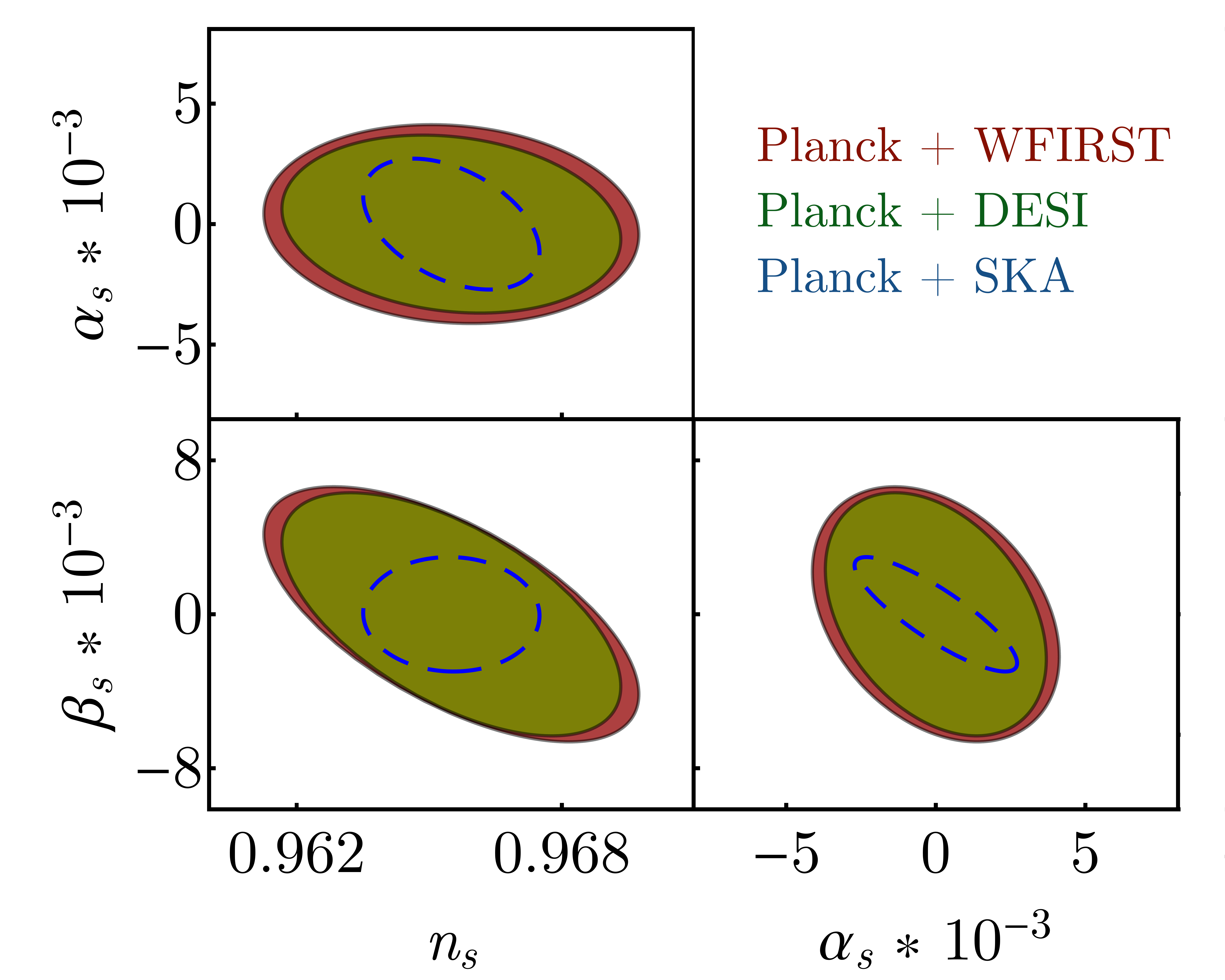}
	\caption{$1-\sigma$ confidence ellipses between $n_s$, $\alpha_s$, and $\beta_s$ for a WFIRST-like experiment in red, for a DESI-like experiment in brown, as well as for a futuristic SKA in blue dashed.
	For all these surveys we add the information from Planck and marginalize over the bias amplitude, as well as the whole set of $\Lambda$CDM parameters $ +$ $ \alpha_s$ and $\beta_s$.
	}
	\label{fig:LSS}
\end{figure*}

We show the confidence ellipses for a DESI-like survey, as well as for a SKA-like survey, in Fig.~\ref{fig:LSS}, where we also include the information from Planck. It is evident from the Figure that both the running $\alpha_s$ and the tilt $n_s$ are degenerate with the second running $\beta_s$ in these measurements.

\subsection{21-cm}
\label{sect:21cm}

The CMB has allowed us to very precisely measure a narrow band of the universe at redshift $z=1100$, 
whereas galaxy surveys like the ones described above map the 
local universe up to $z\sim O(1)$. The region between these two probes has very valuable cosmological information,
which can be probed with the 21-cm hydrogen line \cite{astro-ph/0010468,astro-ph/0312134}. 
In this section we will perform an order-of-magnitude forecast of the capabilities of different experiments observing the 
21-cm line.
This line is emitted by neutral hydrogen when undergoing the hyperfine transition from its triplet to its singlet ground state, and the high neutral-hydrogen
cosmic abundance makes it a potentially powerful cosmological tool.
We can divide the region observable with the 21-cm line into two eras, the epoch of reionization (EoR), from $z\sim 6$ to  11, where hydrogen was heated up by astrophysical processes, and emitted in the 21-cm line; 
and the dark ages, ranging from $z\sim30$ to $z\sim 100$, where the hydrogen cooled adiabatically and resonantly absorbed photons from the CMB at the 21-cm-line frequency.

Several experiments have started observing the EoR, among them the MWA\footnote{http://www.mwatelescope.org/}, and PAPER\footnote{http://eor.berkeley.edu/}, which have obtained
upper limits on the 21-cm power spectrum at $z\sim 7$ \cite{0904.2334,1304.4991,1212.5151,1608.06281}. 
Nonetheless, this effort will yield accurate measurements of the neutral-hydrogen distribution during the EoR, 
especially as HERA\footnote{http://reionization.org/} is built and starts taking data \cite{1606.07473}. 
A very thorough forecast of the capabilities of different experiments
was realized in Refs.~\cite{0802.1710,1501.04291}.
Here we will perform an order-of-magnitude forecast including the second running $\beta_s$, 
albeit with a much more simplistic model of the EoR. 
Our simple approach shows that a measurement of $\beta_s$
is orders of magnitude away of all the EoR experiments, so we believe 
it is not necessary to include additional modeling of the nuisance parameters during this era.

In order to reach scales beyond $k\sim 1$ Mpc$^{-1}$ one could observe at the dark ages, where the
matter distribution of the universe remains linear down to much smaller scales \cite{astro-ph/0312134}, 
enabling access to a tremendous wealth of cosmological information. 
This era, however, will be extremely difficult to probe,
due to the increase in Galactic synchrotron emission at low frequencies, as well as the atmospheric absorption of radio frequencies.
LOFAR\footnote{http://www.lofar.org/} will access the edge of the frequency range 
required for this task, although it is not likely to reach 
enough sensitivity to observe primordial fluctuations \cite{1305.3550}.
Instead, this will require building an interferometric array on the far side 
of the moon \cite{0902.0493,1106.5194}.
The proposed DARE\footnote{http://lunar.colorado.edu/dare/} satellite will serve as a stepping stone to explore the end of the dark ages, although
to constrain the runnings to any significant level one 
will need a large moon-based interferometer, which we will model as different FFTT long-baseline arrays.

\subsubsection{Formalism}

For a review of the physics of the 21-cm line, and its temperature fluctuations, see
for example Refs.~\cite{astro-ph/0608032,1109.6012}. 
At any point
in space we can calculate the difference between the 21-cm temperature $T_{21}(\mathbf x)$ and the average temperature $\overline T_{21} (z)$ at
that redshift as $\Delta T_{21} (\bf x)$, and  $\Delta T_{21} (\bf k)$ is its Fourier transform.
Following the notation in
Ref.~\cite{1506.04152} we write the two-point function of 21-cm temperature fluctuations at redshift $z$ as
\be
\VEV{\Delta T_{21} (\mathbf k) \Delta T_{21} (\mathbf k')}\equiv P_{21}(\mathbf k,z)  (2\pi)^3 \delta_D{(\bf k-k')},
\ee
with
\be
P_{21}(\mathbf k,z)= \left [ {\cal A}(z) + \overline {T}_{21} (z) \mu^2\right]^2 P_{\rm HI}(k,z),
\label{eq:P21}
\ee
where ${\cal A}(z) = d T_{21}/d \delta_b$ is a known function of $z$ (see Refs.~\cite{astro-ph/0401206,1506.04152} for example), $\mu\equiv k_{||}/k$ is the cosine between the line-of-sight $k_{||}$ and $k$, and $P_{\rm HI}$ is the power spectrum of the neutral-hydrogen density perturbations, which equals
$P_{\delta}$ in the scales of interest.

Given an antenna array with a baseline $D_{\rm base}$ uniformly covered up to a fraction $f_{\rm cover}\leq 1$, observing for a time $t_o$, we can write the instrumental-noise power spectrum in $k$-space as \cite{astro-ph/0311514,0805.4414}
\be
P^N_{21} (z) = \dfrac{\pi T_{\rm sys}^2}{t_o f_{\rm cover}^2} \chi^2(z) y_\nu(z) \dfrac{\lambda^2(z)}{D_{\rm base}^2},
\ee
where $\lambda(z)$ is the 21-cm transition wavelength at redshift $z$, $y_\nu(z)= 18.5\,\sqrt{(1+z)/10}$ Mpc/MHz is the conversion function from frequency $\nu$ to $k_{||}$, 
and the system temperature $T_{\rm sys}$ is dominated by the galactic synchrotron emission, parametrized as \cite{0802.1525}
\be
T_{\rm sys} = 180 \, {\rm K} \times \left( \dfrac{\nu}{180\,\rm MHz}\right)^{-2.6}.
\label{eq:Tsys} 
\ee


We take a FFTT-like experiment \cite{0805.4414,0909.0001}, with $f_{\rm cover}=1$ and a variable baseline $D_{\rm base}$, observing 2$\pi$ steradians of the sky (so $f_{\rm sky}=0.5$).
For these experiments
we will increase the pivot scale to $k_*=0.1$ Mpc$^{-1}$,
to help break the degeneracy between $\alpha_s$ and $\beta_s$
arising from the augmented observable $k$ range. 
When combining the constraints from 21-cm measurements with the CMB, 
however, we will use the results at $k_*=0.05$ Mpc$^{-1}$ for
compatibility.

The baseline of each array will determine the maximum perpendicular wavenumber it can observe, calculated as
\be
k_\perp^{\rm max} = \dfrac{2\pi \,D_{\rm base}}{\chi(z) \lambda(z)} \approx \dfrac{2\, \rm Mpc^{-1}}{(1+z) + 1.1\, \sqrt{1+z}} \times \dfrac{D_{\rm base}}{\rm km},
\ee
although during the EoR we do not go beyond the non-linear scale $k_{\rm NL}$, which we set at 1 Mpc$^{-1}$ \cite{0802.1710}. For simplicity we choose a matching line-of-sight resolution for the FFTT experiments to have a single maximum $k_{\rm max}$ that will vary with $D_{\rm base}$, although in practice line-of-sight resolution might be easier to achieve through finer frequency binning. 
We also assume that astrophysical foregrounds will cut off line-of-sight wavenumbers
smaller than \cite{astro-ph/0510027}
\be
k_{||}^{\rm min} \approx \dfrac{2 \pi}{y_\nu\, \Delta \nu},
\ee
where $\Delta \nu$ is the total bandwidth probed by our experiment, and is $\Delta\nu\approx 30$ MHz for the EoR ($z=8-10$) and $\Delta\nu\approx 50$ MHz for the dark ages ($z=20-100$),
which translates into a minimum wavenumber $k_{||}^{\rm min} \approx 10^{-2}$ Mpc$^{-1}$ for both eras. Our results do not depend sensitively on this cutoff, so we set both $k_{||}^{\rm min}$ and $k_{\perp}^{\rm min}$ to $k_{\rm min}=10^{-2}$.

Similarly to the CMB, the 21-cm power spectrum $P_{21}(k)$ depends on the primordial parameters $n_s$, $A_s$, $\alpha_s$, and $\beta_s$ 
only through the matter power spectrum $P_{\delta}(k)$. Unfortunately, the redshift functions $\mathcal A(z)$
and $T_{21}(z)$ depend on the rest of $\Lambda$CDM parameters. 
For simplicity we will perform an order-of-magnitude forecast of the 
errors only on the primordial parameters, ignoring $\omega_b$, $\omega_c$, $H_0$, and 
the reionization parameters. This is a vast oversimplification, 
adopted because the primordial parameters
should be moderately decoupled from the rest of $\Lambda$CDM.
For a more complete discussion see Ref.~\cite{0802.1710}.

To forecast we separate the available comoving volume in redshift bins, chosen to be small enough that all redshift-dependent parameters are constant within each bin.
We then compute the Fisher matrix for one of these slices, 
centered at redshift $z_i$, as \cite{astro-ph/9706198}
\be
F_{a b}^{(i)} = \dfrac { f_{\rm sky}} 2 \dfrac{{\rm Vol}_i}{(2\pi)^3} \int_{k_{\rm min}}^{k_{\rm max}} \!\!\! \!\!\! \!\!\!  \! \mathrm d k (2\pi k^2) \int_{-1}^{1} \!\!\!\mathrm d \mu  \dfrac{\dfrac{\partial P_{21}(\mathbf k,z)}{\partial p_a}\dfrac{\partial P_{21}(\mathbf k,z)}{\partial p_b}}{\left [ P_{21}(\mathbf k,z) + P_{21}^N(z)\right]^2},
\ee
where ${\rm Vol}_i$ is the comoving volume of the slice, and $p_a=\{A_s,n_s,\alpha_s,\beta_s\}$. 
We will then incorporate the information from all the redshift bins by just adding the Fisher matrices, i.e.,
\be
F_{a b} = \sum_i F_{a b}^{(i)}.
\label{eq:Fisher21}
\ee

\subsubsection{EoR Results}

During the EoR the spin temperature $T_S$ is coupled to the gas temperature due to the Wouthuysen-Field effect \cite{Wout,Field,astro-ph/0507102}. Heating of the gas due to stellar formation causes the spin temperature to rise above the CMB temperature $T_\gamma$, so the 21-cm line should be visible in emission. During this era we can write the factors in Eq.~\eqref{eq:P21} as
\be
\mathcal A(z) = \overline T_{21} (z) = 27.3\, {\rm mK}\,\times \overline x_H \dfrac{T_S - T_{\gamma}}{T_S} \left ( \dfrac{1+z}{10} \right)^{1/2},
\label{eq:T21bar}
\ee
where $\overline x_H$ is the mean neutral-hydrogen fraction, and we can drop the temperature factor since $T_S\gg T_\gamma$ \cite{0802.1710}. We adopt the simplest model of reionization, in which the free-electron fraction is parametrized as \cite{0804.3865}
\be
x_e(z) = \dfrac{1+f_{\rm He}}{2} \left[1+\tanh \left( \dfrac{f(z_*) - f(z)}{\Delta f}\right)\right],
\ee
where $f_{\rm He}\approx 0.082$ is the Helium fraction, $f(z)\equiv (1+z)^{3/2}$, and $\Delta f = 3(1+z_*)\Delta z/2$.
The best-fit redshift for this model is $z_*=8.8$, with a width $\Delta z=0.5$ \cite{1605.03507}, although
this last parameter is poorly constrained, as it has little effect on the CMB as long as it is small.

The free-electron fraction $x_e$ will also fluctuate, giving rise to a power spectrum $P_{x}$, as well as to a cross-spectrum with density fluctuations $P_{\delta x}$ \cite{astro-ph/0311514,astro-ph/0604177}. However, before reionization started there was likely a regime in which $T_S\gg T_\gamma$, producing 21-cm emission, albeit with a mostly neutral medium, where we can neglect the $x_e$ perturbations \cite{0802.1710}. We will only forecast in this regime, which we will assume extends from $z = 8$ to $10$. Reionization could have lasted significantly longer \cite{1605.03507}, providing additional observing volume. On the other hand, complications arising from a 
nonvanishing $P_{x}$ and $P_{\delta x}$ will make our simple forecast too optimistic, so it should be treated as an order-of-magnitude estimate.

\begin{figure}[htbp!]
	\includegraphics[width=0.4\textwidth]{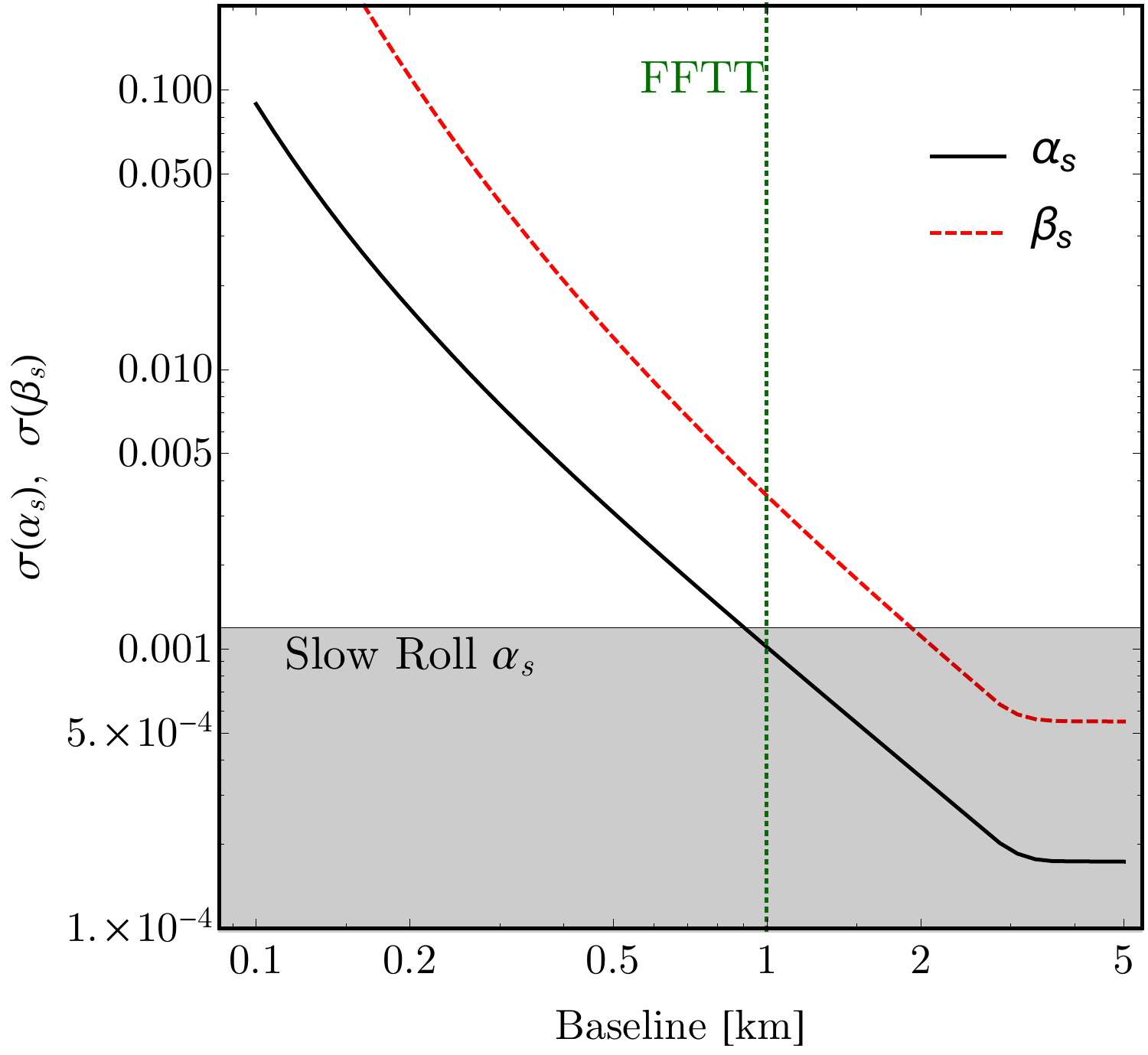}
	\includegraphics[width=0.383\textwidth]{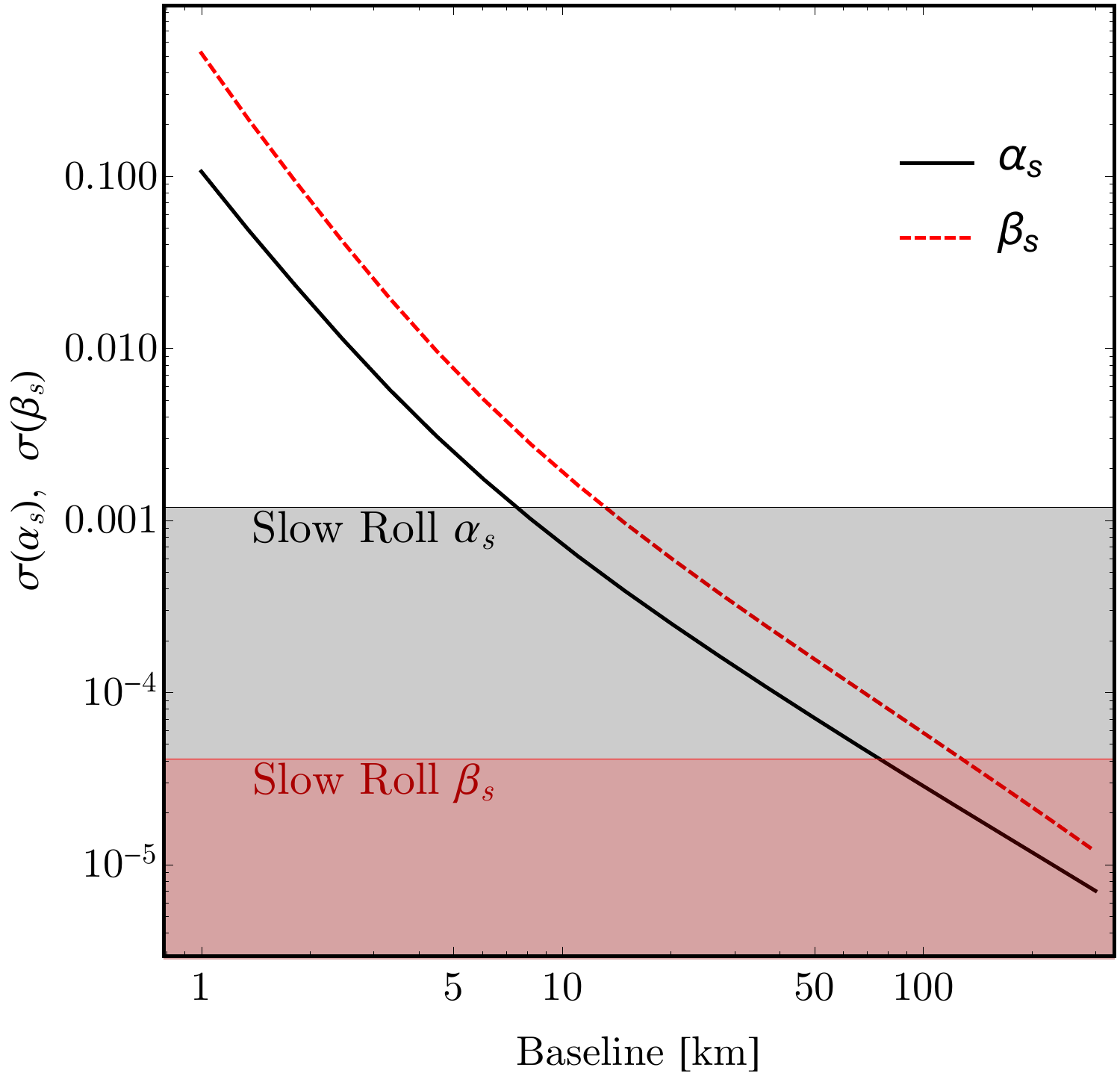}
	\caption{1$-\sigma$ uncertainties in $\alpha_s$ and $\beta_s$ for an experiment measuring the epoch of reonization (\emph{left}) and the dark ages (\emph {right}) with a dense core ($f_{\rm cover}=1$) extending for a baseline $D_{\rm base}$ in km. We show the slow-roll-inflation prediction for $\alpha_s$ in gray and for $\beta_s$ in red, and in dashed green we show the specifications for the proposed FFTT.
	In this figure we have marginalized over $A_s$ and $n_s$ to calculate $\sigma(\alpha_s)$, and also over $\alpha_s$ to compute $\sigma(\beta_s)$, and not marginalized over nuisance parameters. Here we have chosen a pivot scale of $k_*=0.1$ Mpc$^{-1}$.
	}
	\label{fig:EoR}
\end{figure}

Using the Fisher matrix in Eq.~\eqref{eq:Fisher21} we can compute the 1$-\sigma$ uncertainties in the primordial parameters, which we show in Table~\ref{tab:21cm}.
Moreover, in Fig.~\ref{fig:EoR} we show the uncertainties in
$\alpha_s$ and $\beta_s$ as a function of the baseline of a fully-covered FFTT experiment. 
The sensitivity curve flattens at $D_{\rm base}\approx 3$ km, corresponding to the
non-linear scale $k_{\rm NL}$, beyond which we assume no information can be extracted.
The proposed FFTT, with a 1-km baseline, should be able to detect the $\alpha_s$ from 
inflation, and increased baselines will provide improved sensitivity. 
Nonetheless, the smallest 
$\beta_s$ that can be measured with the FFTT is still two orders of magnitude 
larger than the inflationary prediction. 
We label this case as FFTT, and show its predicted confidence ellipses between $n_s$, $\alpha_s$, and $\beta_s$ in Fig.~\ref{fig:LSS}, and the marginalized constraints in Table~\ref{tab:21cm}. 
We again emphasize that these forecasted uncertainties are good to an order-of-magnitude level,
as we have neglected correlations with the non-primordial $\Lambda$CDM parameters.
Notice also that in Ref.~\cite{1501.04291} it was shown that HERA can improve the Planck constraint on $\alpha_s$ by $\sim 20\%$, lowering it to $\sigma(\alpha_s)=0.0036$, so it is unlikely it will improve the S4 results significantly.

\subsubsection{Dark Ages Results}

During the dark ages the spin temperature is coupled to the gas temperature through collisions \cite{astro-ph/0312134}. This makes $\overline T_{21}$ negative, causing absorption of CMB photons at radio frequencies. The redshift range in which $\overline T_{21},\mathcal A < 0$ is $z \sim 20-200$. 
During this range $\overline T_{21}$, as well as $\mathcal A(z)$, including perturbations to gas temperature, can be found in Ref.~\cite{1506.04152}.
We note that the $z\sim 20$ range might be contaminated by astrophysical effects, such as heating due to star formation \cite{astro-ph/0304131} or miniquasars \cite{astro-ph/0506712}. In any case our results are not altered dramatically by changes in the starting redshift.

With the same procedure as for the EoR, but over the redshift range $z = 20-100$, we can compute the uncertainty in $\alpha_s$, when marginalizing over $A_s$ and $n_s$, for a FFTT-like experiment. Likewise for $\beta_s$, marginalizing in this case over $\alpha_s$ as well. We show these errors in Fig.~\ref{fig:EoR}, where it can be seen that an experiment with a baseline of  $D_{\rm base}\sim 5$ km could confirm the slow-roll prediction for $\alpha_s$, whereas to detect $\beta_s\sim(1-n_s)^3$ one would need $D_{\rm base}= O(100)$ km. With the goal of detecting a non-vanishing $\beta_s$ we propose a 300-km perfectly covered array, which we just label FFTT$_{300}$. We show the results for this very-futuristic array in Table~\ref{tab:21cm}.

\begin{table*}[hbtp!]
	{\small
		\begin{tabular}{|l|c|c|c|c|}
			\hline
			Array & $\sigma\left(A_{s}\right)$ & $\sigma\left(n_{s}\right)$ & $\sigma\left(\alpha_s\right)$ & $\sigma\left(\beta_s\right)$ \\
			\hline\hline
			FFTT & $6.5 \times 10^{-13}$ & $3.3 \times 10^{-4}$ & 0.0010 & 0.0035
			\\ FFTT$_{300}$ & $1.4 \times 10^{-15}$ & $4.1\times 10^{-6}$ & $7.0\times 10^{-6}$ & $1.2\times 10^{-5}$
			\\ \hline
		\end{tabular}
	}
	\caption{$1-\sigma$ uncertainties in the scalar amplitude $A_s$, tilt $n_s$, running $\alpha_s$, and second running $\beta_s$ for different 21-cm arrays. For $A_s$, $n_s$, and $\alpha_s$ we only marginalize over $A_s$ and $n_s$, whereas for $\beta_s$ we marginalize over $\alpha_s$ as well, and we do not marginalize over nuisance parameters. Here we have increased the pivot scale to $k_*=0.1$ Mpc$^{-1}$.
	}
	\label{tab:21cm}
\end{table*}

\subsection{Combined constraints}
\label{sect:combined}

The strength of the Fisher-matrix approach we follow is that we can easily add the information from
different experiments by summing their Fisher matrices. We take as a our baseline case the proposed S4 CMB
experiment, and in Table~\ref{tab:minrunning} we show the minimum $\alpha_s$ and $\beta_s$ observable at 1$-\sigma$
when adding different combinations of experiments to the S4 CMB. The results for 
CMB and galaxy surveys marginalize over all $\Lambda$CDM parameters and the bias $b$ for $\alpha_s$, and we include $\alpha_s$ in the marginalization
for $\beta_s$. 
These results show that only very-futuristic experiments will
be able to probe the dynamics of inflation, with the S4+SKA reaching a sensitivity of $\sigma(\alpha_s) \approx 10^{-3} \lesssim (1-n_s)^2$
Interestingly, in all cases $\sigma(\beta_s) \sim 2 \,\sigma(\alpha_s)$, which attests to the difficulty of determining the second running.

	\begin{table}[hbtp!]
		\begin{tabular}{| l | c | c |}
			\hline
			Experiment &  $\sigma(\alpha_s)$ & $\sigma(\beta_s)$   \\             
			\hline
			\hline
			S4 CMB & 0.0025 & 0.0045 \\
			S4+WFIRST & 0.0020 & 0.0035 \\
			S4+DESI & 0.0018 & 0.0034 \\
			S4+SKA & 8.5 $\times 10^{-4}$ & 0.0019 \\
			\hline
		\end{tabular} \qquad\qquad\qquad
		\caption{$1-\sigma$ uncertainties on the running $\alpha_s$ and the second running $\beta_s$ for the S4 CMB experiment plus different proposed galaxy surveys.
		We marginalize over the six $\Lambda$CDM parameters (plus the bias amplitude $b$ for the galaxy surveys) when computing $\sigma(\alpha_s)$ and over $\alpha_s$ as well for $\sigma(\beta_s)$.
		}
		\label{tab:minrunning}
	\end{table}

\begin{figure}[htbp!]
	\hspace*{-1.5cm}
	\includegraphics[width=0.6\textwidth]{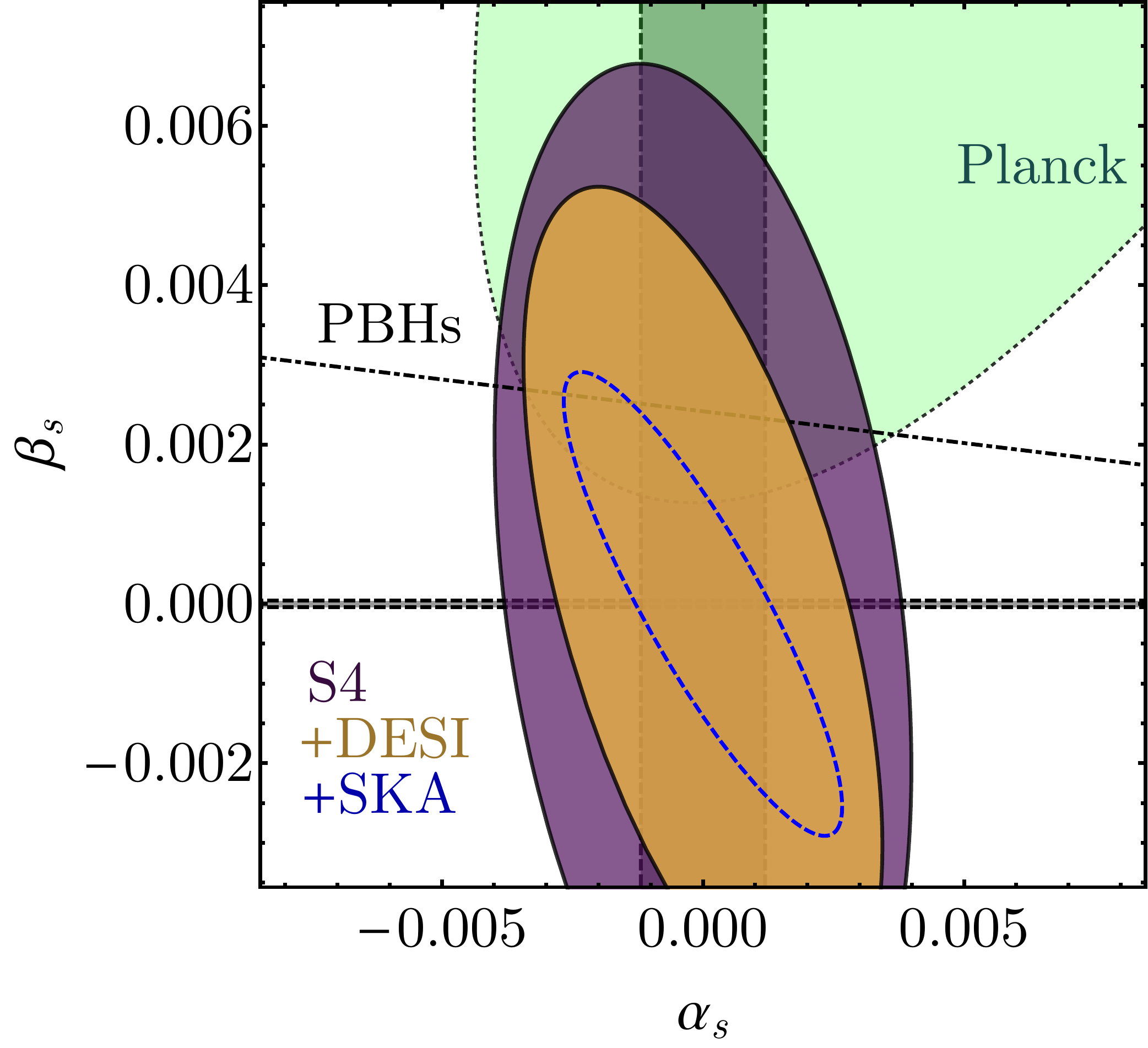}
	\caption{68\% confidence ellipses in the $\alpha_s-\beta_s$ plane for the S4 CMB experiment (purple), S4+DESI (yellow), and S4+SKA (blue). We show the current Planck ellipse from Ref.~\cite{1605.00209} in green. In gray we plot the range predicted by slow-roll single-field inflation. The region above the dash-dotted black line could produce PBHs with masses $M_{\rm pbh}>10^{15}$ gr, if extrapolated to the smallest scales.}
	\label{fig:total}
\end{figure}

In Fig.~\ref{fig:total} we plot the $1-\sigma$ confidence ellipses in the $\alpha_s-\beta_s$  plane for the S4 CMB experiment, as well as
the combination of S4+DESI, and S4+SKA. A representation of the current Planck 68\% C.L. region is also shown (from Ref.~\cite{1605.00209}), which displays a slight preference for a non-zero $\beta_s$.
We draw a line in the $\alpha_s-\beta_s$ plane above which PBHs with masses larger than $10^{15}$ gr could be formed, 
calculated as in Sec.~\ref{sect:motivation}. 
Comparing the ellipses in Fig.~\ref{fig:total} with the predictions
from slow-roll inflation it is clear that departures from slow-roll behavior should be detectable
in $\alpha_s$, although not in $\beta_s$ unless they are very drastic.

\section{Conclusions}
\label{sect:conclusions}

In this paper we have studied how well different future observables will be able to measure the 
runnings of the scalar power spectrum. A summary of our results is in Table~\ref{tab:minrunning}.
Perhaps the most promising of these future probes is
the S4 CMB experiment, proposed in Ref.~\cite{1610.02743}, which will be able to measure the
$\Lambda$CDM parameters to astounding precision (see Table~\ref{tab:S4}). 
This S4 CMB 
experiment will probe the runnings to a precision $\sigma(\alpha_s)= 0.0025$ and $\sigma(\beta_s)= 0.0045$, insufficient to 
detect the single-field slow-roll inflation prediction, although
enough to measure significant departures from it. 

We added to the S4 CMB results the information from upcoming galaxy surveys, 
such as WFIRST and DESI. We find that these surveys will marginally improve the S4 measurements, reducing the error bars by at most 30\%.
However, more futuristic surveys, such as a billion-object SKA, will add enough information to half the S4 CMB uncertainties,
reaching enough sensitivity to detect $\alpha_s\sim10^{-3}$, as predicted by slow-roll inflation.
Additionally, these measurements will allow us to falsify the model for PBH production proposed in Ref.~\cite{1607.06077}
with the improved $\beta_s$ accuracy.

To detect the slow-roll prediction for $\beta_s$, however, 
it will be necessary to probe many more modes, which might be possible with the 21-cm line. 
The proposed FFTT, with a 1-km baseline, will not be sensitive enough to detect $\beta_s \sim 10^{-5}$.
Only a very-futuristic and lunar-based FFTT observing at the dark ages,
with a 300-km baseline, will reach enough modes to be guaranteed a measurement of both $\alpha_s$ and $\beta_s$ from inflation.

To summarize, within the next few decades the uncertainties in the runnings $\alpha_s$ and $\beta_s$
will decrease by a significant factor, as new cosmological experiments are developed and their data is analyzed. This will
allow us to very-precisely characterize the dynamics of inflation, and test deviations from the standard slow-roll scenario. 
Such a measurement will be invaluable for characterizing the inflationary potential beyond the first-order slow-roll approximation,
opening a window into the first moments of the Universe.

	\begin{acknowledgments}
		We thank Simeon Bird and Cora Dvorkin for useful discussions. 
		This work is supported at Johns Hopkins University by the Simons Foundation, NSF grant PHY-1214000, and NASA
		ATP grant NNX15AB18G. 
		AR has received funding
		from the People Programme (Marie Curie Actions) of the European Union H2020 Programme under REA grant agreement number 706896 (COSMOFLAGS). Funding for this work was partially provided by the Spanish MINECO under MDM-2014-0369 of ICCUB (Unidad de Excelencia ``Mar\'ia de Maeztu'').
		The work of JS was supported by ERC Project No. 267117 (DARK) hosted by the Pierre et Marie Curie University-Paris VI.
		JBM also thanks the Institut d'Astrophysique de Paris and the Instituto de F\'isica Te\'orica de Madrid, where part of this work was carried out.
	\end{acknowledgments}

\end{document}